\documentclass[a4paper,11pt]{article}
\usepackage{pos}

\title{Hyperonic degrees of freedom in binary neutron star mergers}
\author*[a,b]{Laura Tolos}
\author[c]{Hristijan Kochankovski}
\author[c]{Angels Ramos}
\author[d,e]{Georgios Lioutas}
\author[d,f]{Sebastian Blacker}
\author[d,g]{Andreas Bauswein}

\affiliation[a]{Institute of Space Sciences (ICE, CSIC), Campus UAB, Carrer de Can Magrans, 08193 Barcelona, Spain}
\affiliation[b]{Institut d'Estudis Espacials de Catalunya (IEEC), 08860 Castelldefels (Barcelona), Spain}
\affiliation[c]{Departament de F\'{\i}sica Qu\`antica i Astrof\'{\i}sica and Institut de Ci\`encies del Cosmos, Universitat de Barcelona, Mart\'i i Franqu\`es 1, 08028, Barcelona, Spain}
\affiliation[d]{GSI Helmholtzzentrum f\"ur Schwerionenforschung, Planckstra{\ss}e 1, 64291 Darmstadt, Germany}
\affiliation[e]{Heidelberger Institut für Theoretische Studien (HITS), Schloss-Wolfsbrunnenweg 35, 69118 Heidelberg, Germany}
\affiliation[f]{Institut f\"ur Kernphysik, Technische Universit\"at Darmstadt, 64289 Darmstadt, Germany}
\affiliation[g]{Helmholtz Research Academy Hesse for FAIR (HFHF), Campus Darmstadt, 64291 Darmstadt, Germany}

\emailAdd{tolos@ice.csic.es}

\abstract{We analyze the influence of hyperons in binary neutron star mergers considering several different equations of state that include hyperons. By running a large set of simulations, we study the impact of the thermally produced hyperons on the gravitational-wave spectral features, the temperature evolution of the remnant, the mass ejecta and the threshold mass for prompt collapse to a black hole. Models with hyperons tend to stand out in the relation between the dominant postmerger gravitational-wave frequency and the tidal deformability of massive stars.  Moreover, the averaged temperature of the remnant is reduced for hyperonic models. The mass ejection of the mergers is tentatively enhanced when  hyperons are present in comparison to nucleonic EoSs leading to similar stellar properties of cold neutron stars, whereas the threshold mass for prompt black-hole formation is reduced by about 0.05~$M_\odot$ compared to the nucleonic models.
}

\FullConference{The 21st International Conference on Hadron Spectroscopy and Structure (HADRON2025)\\
 27 - 31 March, 2025\\
Osaka University, Japan\\}

\begin{document}
\maketitle

\section{Introduction}\label{sec:intro}

Over decades the composition of neutron stars (NSs) has been a matter of debate. Among the different forms of matter, hyperons in the core of NSs have been extensively studied (see for example the recent reviews of Refs.~\cite{Chatterjee:2015pua,Tolos:2020aln}). The appearance of hyperons has been mainly discussed in the context of  the so-called hyperon puzzle, i.e.~the tension between the softening of the equation of state (EoS) due to the appearance of hyperons and the observations of NSs with $M \gtrsim 2M_{\odot}$ \cite{Demorest2010ShapiroStar,Antoniadis:2013pzd,Fonseca2016,Cromartie2020RelativisticPulsar,Romani:2022jhd}.

Distinguishing the presence of hyperons by analysing the masses and radii of  cold NS is however not an easy task. For reference, we mention Ref.~\cite{Bauswein:2025dfg}, where the identification of hyperons through the negative curvature of the mass-radius relation  has been put forward. 

A new venue for detection of hyperons has been proposed recently \cite{Sekiguchi:2011mc,Radice:2016rys,Blacker:2023opp,Kochankovski:2025lqc} in the context of binary neutron star (BNS) mergers~\cite{LIGOScientific:2017vwq,Abbott2017em,Abbott2018GW170817:State,Abbott2019}. While early works have studied a very limited set of hyperonic models ~\cite{Sekiguchi:2011mc,Radice:2016rys}, in Refs.~\cite{Blacker:2023opp,Kochankovski:2025lqc}  a large sample of hyperonic EoSs has been considered in the simulations of BNS mergers, showing that the thermal behavior of the EoSs can serve as a potential discriminator between hyperonic and purely nucleonic matter in  BNS mergers. In Refs.~\cite{Blacker:2023opp,Kochankovski:2025lqc} the impact of the thermally produced hyperons on BNS has been analysed, such as in the gravitational-wave (GW) spectral features, the averaged temperature of the remnant, the mass ejection and the threshold binary mass for prompt black-hole formation. In the present paper we review these latest results.

\section{Thermal behaviour of hyperonic models} \label{sec:EoS}

\begin{figure*}[htbp]
    \centering
        \includegraphics[width=0.45\textwidth]{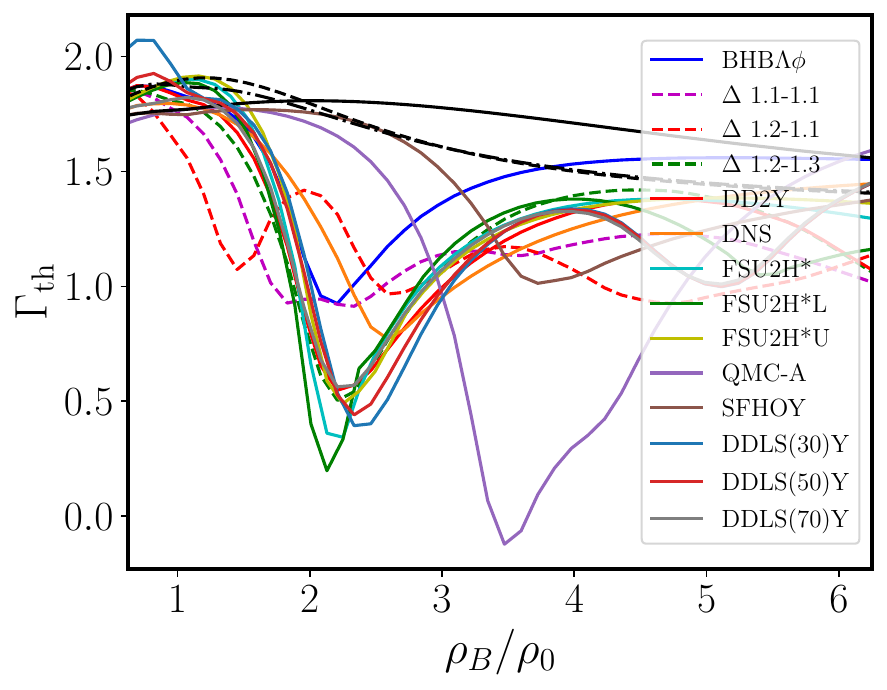}
        \includegraphics[width=0.45\textwidth]{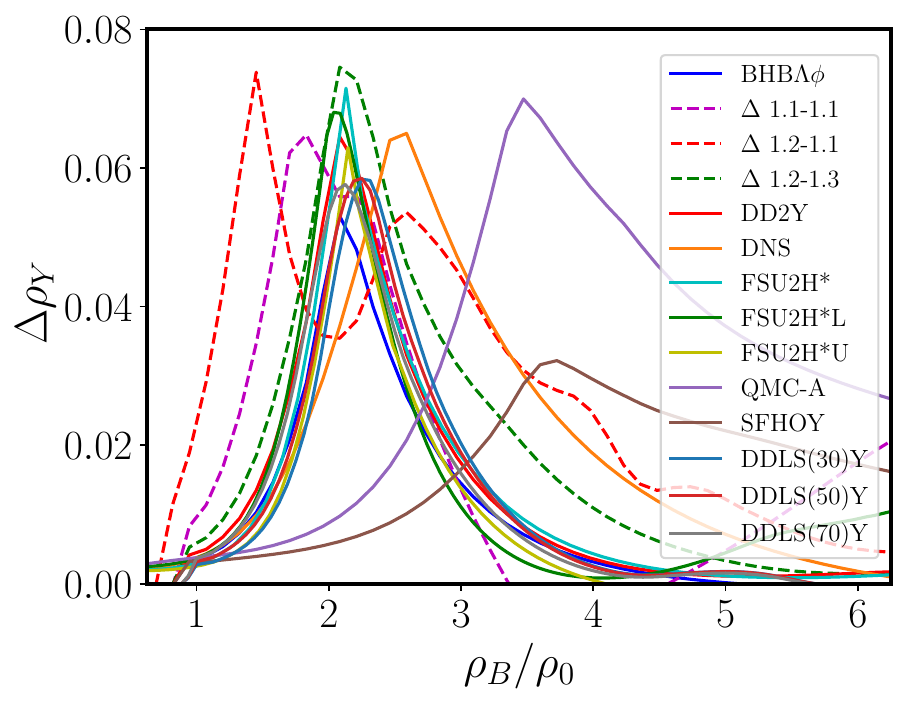}
    \caption{$\Gamma_{\mathrm{th}}$ (left plot) and $\Delta \rho_Y$ (right plot) as functions of density at  $Y_Q = 0.1$ and  $T = 25$ MeV for different hyperonic (colored curves) and nucleonic (black curves) models. The three nucleonic models are: solid line for SFHO, dashed line for FSU2R and dash-dotted line for DD2. Plots from Ref.~\cite{Kochankovski:2025lqc}.}
    \label{fig:thermal}
\end{figure*}

We present a review on the investigation of the thermal behaviour of hyperons in BNS mergers. By using a large set of hyperonic and nucleonic models (see Ref.~\cite{Kochankovski:2025lqc} for more details on the models), we quantify the  thermal behavior of the thermal energy density ($\epsilon_{\mathrm{th}}$) and thermal pressure ($P_{\mathrm{th}}$) for nucleonic and hyperonic models by means of the thermal index: 
$\Gamma_\mathrm{th} (\rho_B,Y_Q,T) = 1+ P_{\mathrm{th}}/\epsilon_{\mathrm{th}}$, 
where $\rho_B$ is the baryonic density, $Y_Q$ the charge fraction and $T$ the temperature. While the thermal index of nucleonic models shows a mild dependence on the density, the behavior of the thermal index for hyperonic (or hyperonic and $\Delta$-baryonic) models is very different (see left plot of Fig.~\ref{fig:thermal}). This is due to the strong reduction of the thermal pressure leading to a significant drop of the thermal index  \cite{Kochankovski2022}, when hyperons are thermally produced. This is clearly seen  in the right plot of the same figure by displaying the normalized hyperon excess as a function of density,  defined as the difference between the hyperon abundance at a given temperature $T$ and the hyperon abundance at $T=0$ for the same density and charge fraction, normalized to the total baryon density: $\Delta \rho_Y =\sum_{Y_i} (\rho_{Y_i}(T,\rho_B,Y_Q) - \rho_{Y_i}(0,\rho_B,Y_Q))/\rho_B$, where  $\rho_{Y_i}$ indicates the density of an arbitrary heavy baryon $Y_i$.

\section{BNS merger simulations} \label{sec:Simulations}

The BNS merger simulations are performed with a relativistic smooth particle hydrodynamics (SPH) code, which adopts the conformal flatness condition to solve the Einstein field equations~\cite{1980grg1.conf...23I,1996PhRvD..54.1317W} (see~\cite{Blacker:2023opp} for more details and~\cite{Oechslin:2001km,2007A&A...467..395O,Bauswein:2009im}  for information on the code). We use the code for  1.4-1.4~$M_\odot$ binaries as well as for asymmetric BNSs and heavier BNS mergers. For all calculations we set up the stars without intrinsic spin, whereas neutrinos and magnetic fields are not included.

Two sets of simulations with all purely nucleonic and hyperonic models were conducted. First, we performed simulations employing the full temperature-, density- and composition- dependent EoS tables. Then, for the other set of simulations, we used all EoSs at temperature $T=0$ under neutrinoless $\beta$-equilibrium conditions. These barotropic slices $P_\mathrm{cold}(\rho)$ and $\epsilon_\mathrm{cold}(\rho)$ are supplemented with an approximate thermal treatment, where the thermal pressure $P_\mathrm{th}$ is given by  $P_\mathrm{th}=(\Gamma_\mathrm{th}-1)\epsilon_\mathrm{th}$ with $\epsilon_\mathrm{th}=\epsilon - \epsilon_\mathrm{cold}(\rho)$ (see~\cite{Janka1993,Bauswein:2010dn}). The chosen thermal index is $\Gamma_\mathrm{th}=1.75$, as it reproduces well the thermal behaviour of  purely nucleonic EoSs. The goal is to discriminate between nucleonic and hyperonic matter from their different thermal behavior with high accuracy future measurements.

\section{Dominant post-merger GW frequency}
\label{sec:fpeak}

\begin{figure*}[htbp]
    \centering
\includegraphics[width=0.45\textwidth]{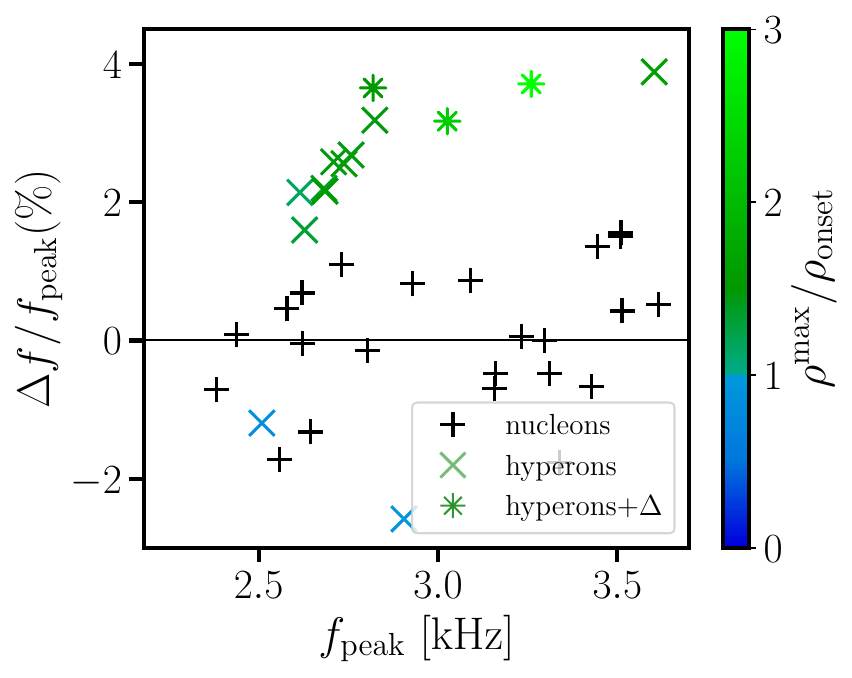}
\includegraphics[width=0.45\textwidth]{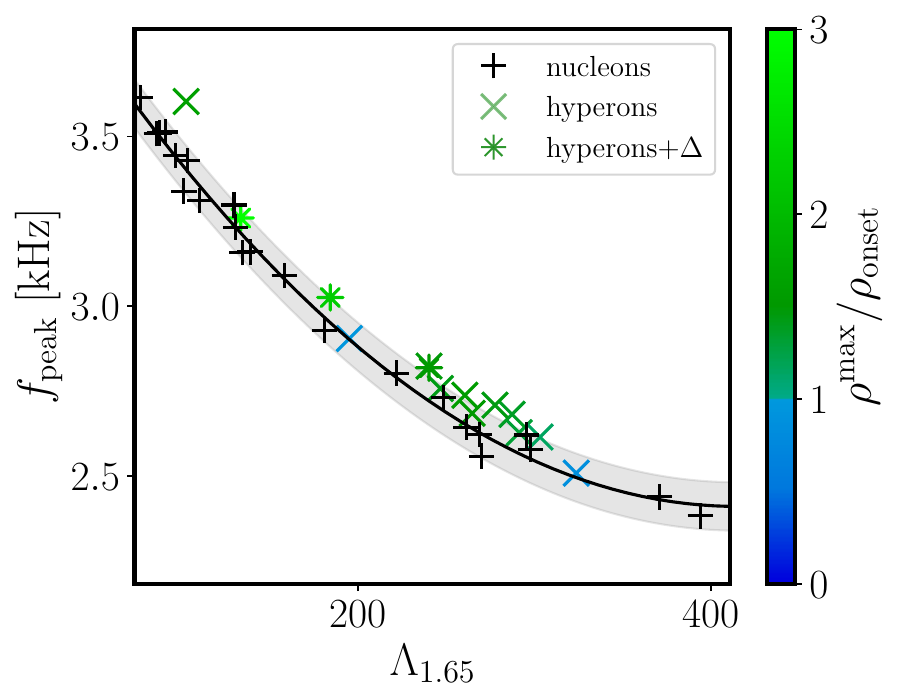}
    \caption{Left plot:  $\Delta f/f_{\mathrm{peak}}$ as a function of $f_{\mathrm{peak}}$. Right plot: $f_{\mathrm{peak}}$ as function of the tidal deformability of a 1.65~$M_\odot$ symmetric BNS merger. The black curve is a least-squares quadratic fit to purely nucleonic models, whereas the gray band shows the maximum residual of purely nucleonic models. Plots  from Ref.~\cite{Kochankovski:2025lqc}.}
    \label{fig:frequency}
\end{figure*}

To assess the influence of the distinctive finite-temperature behaviour of the hyperonic models on BNS observables, we start by studying the dominant post-merger GW frequency. We define $\Delta f = f_\mathrm{peak} - f_\mathrm{peak}^{1.75}$, with $f_\mathrm{peak}$ being the dominant GW frequency from the simulation with the fully temperature dependent EoSs and $f_\mathrm{peak}^{1.75}$ being the dominant frequency from  the run with the zero-temperature beta-equilibrium EoS slice of the same model and the approximate "nucleonic" thermal treatment of  $\Gamma_\mathrm{th}=1.75$. Hence, the quantity $\Delta f$ is a measure of how much a given model deviates from an ``idealized'' nucleonic thermal behavior. 

In Fig.~\ref{fig:frequency}  we show the relative frequency shift as a function of  $f_\mathrm{peak}$ (left plot). Models with hyperons are labeled with crosses, models that also contain $\Delta$ baryons are labeled with asterisks, whereas the purely nucleonic models are indicated with plus signs. The color bar shows the ratio between the maximum rest-mass density in the merger remnant within the first 5 ms after the merger and the density at which heavy baryons start to appear in matter in $\beta$ equilibrium at $T=0$. We find that the purely nucleonic models scatter around a zero relative frequency shift, while the hyperonic models show a systematic shift towards higher values in the range of $2\%-4\%$. If the density in the remnant is low, a frequency shift is not observed for hyperonic models, since the abundance of hyperons (and/or $\Delta$s) is too small. In the right plot of Fig. \ref{fig:frequency} we display the $f_{\mathrm{peak}}$ as a function of the tidal deformability for a 1.65~$M_\odot$ symmetric BNS merger  ($\Lambda_{1.65}$)  for all full EoS models considered in this work.  The  solid line is the quadratic fit to the nucleonic data and the gray band indicates the maximum residual of the nucleonic models with respect to the fit. We show that, for a given tidal deformability, the dominant frequency of most of the models with hyperons (or hyperons and $\Delta$s)  is higher compared to that of the nucleonic models. We note that similar or even larger shifts are obtained for asymmetric binaries or for more massive stars.

\section{Temperature evolution of the remnant}
\label{sec:temp}

\begin{figure*}[htbp]
    \centering
 \includegraphics[width=0.45\textwidth]{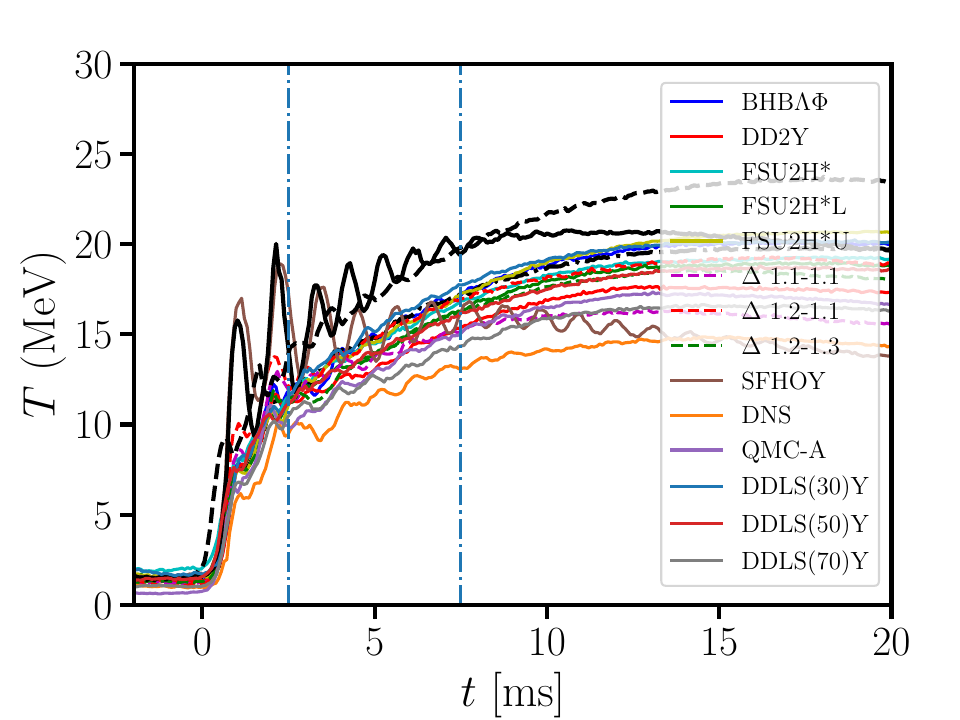}
 \includegraphics[width=0.45\textwidth]{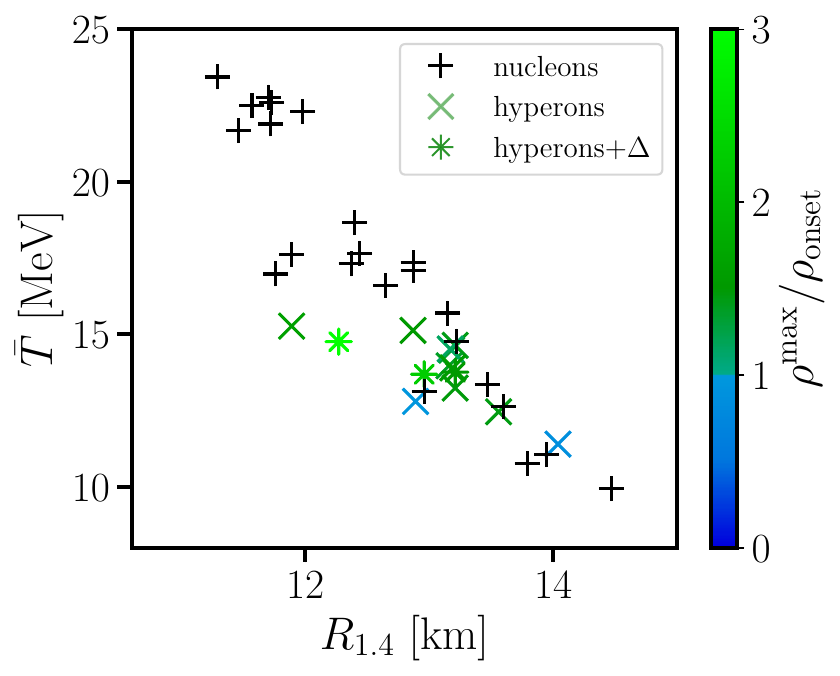} 
\caption{Left plot: Mass averaged temperature ($T$) of the remnant as a function of time ($t$) for hyperonic models. With black lines,  three nucleonic models (DD2 - dotted line, SFHO - solid line, FSU2R - dashed line) are shown. The vertical lines show the time window  to compute the mass and time averaged temperature ($\bar{T}$). Right plot: $\bar{T}$ in the remnant  as a function of the radius of a $1.4M_{\odot}$ NS. Plots from Ref.~\cite{Kochankovski:2025lqc}.}
\label{fig:temp}
\end{figure*}

We now analyse the influence of hyperons on the temperature evolution of the remnant.
In Fig.~\ref{fig:temp} we show the mass averaged temperature  of the remnant as a function of time (left plot). The presence of hyperons reduces the averaged temperature with respect to the case when only nucleons are present (colored curves as compared to the black lines).  This is due to the fact that the specific heat is significantly larger for hyperonic matter than for purely nucleonic matter. Hence, for a given amount of thermal energy, the temperature increase of matter is lower in the hyperonic  models. In the right plot of Fig.~\ref{fig:temp} we display the mass and time averaged temperature ($\bar T$) in the remnant  as function of the radius of a $1.4M_{\odot}$ star.  The averaged  temperature roughly anticorrelates with the NS radius, whereas hyperonic models yield temperatures which are reduced compared to purely nucleonic models with the same radius.

\section{Mass ejection and threshold mass for prompt collapse}
\label{sec:mass-threshold}

\begin{figure*}[htbp]
\centering
\includegraphics[width=0.45\textwidth ]{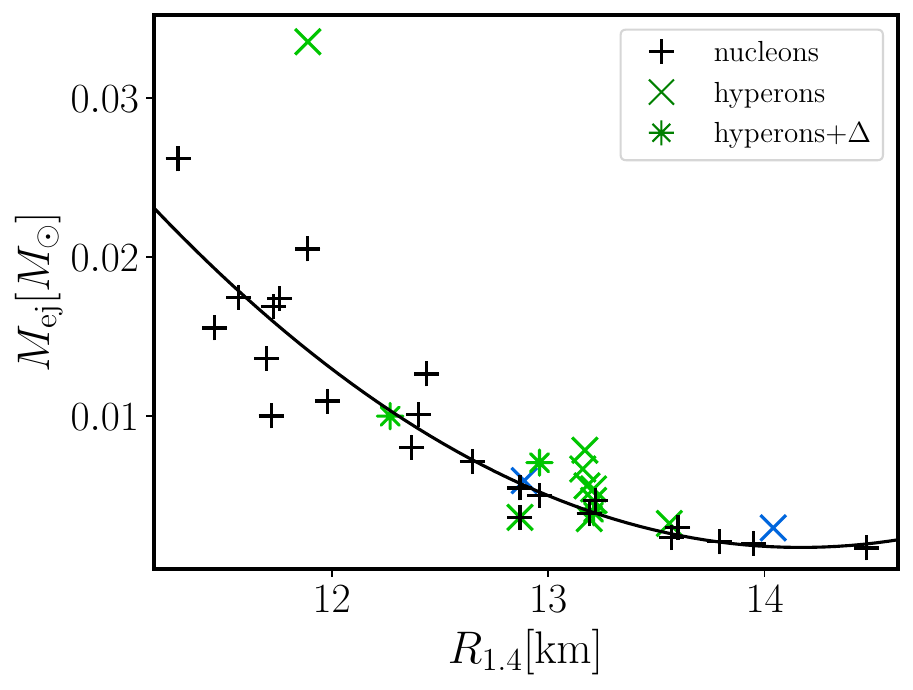}
\includegraphics[width=0.45\textwidth ]{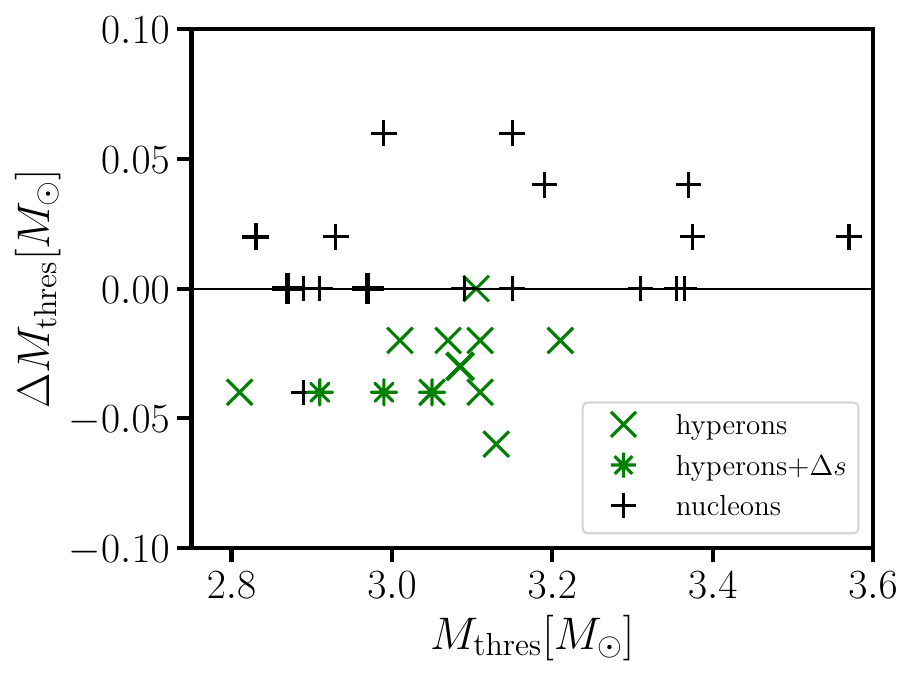}
\caption{Left plot:  $M_{\mathrm{ej}}$ as a function of $R_{1.4}$.  The black curve shows a parabolic fit to the purely nucleonic models. Right plot: The difference $\Delta M_{\mathrm{thres}}=M_\mathrm{thres}-M_{\mathrm{\mathrm{thres}}}^{1.75}$ as function of $M_{\mathrm{thres}}$. Plots from Ref.~\cite{Kochankovski:2025lqc}. } 
\label{fig:mass-threshold}
\end{figure*}

The mass ejection of mergers as a potential signature of the presence of hyperons is shown in the left plot of Fig.~\ref{fig:mass-threshold}. The ejecta mass is of particular interest since it determines the total amount of synthesized elements and the brightness of the electromagnetic emission in the optical and infrared in BNS mergers. In the left plot we show the ejecta mass ($M_{\mathrm{ej}}$) of simulations with 1.4-1.4~$M_\odot$ binaries as a function of the radius ($R_{1.4}$), as an approximate measure for the EoS stiffness, where the amount of unbound material is determined 10~ms after merging. The softer the EoSs (smaller radii) are, the significant higher ejecta masses are obtained. We observe an enhancement due to the appearance of hyperons for the same models, although it is not generic and universal.

It is also possible to identify the correlation between the highest total mass for which the remnant does not collapse promptly to a black hole (on a timescale of about a millisecond after the merging) and the appearance of thermal hyperons. In Fig.~\ref{fig:mass-threshold} (right plot) we compare the threshold mass resulting from the simulations with the fully temperature dependent EoSs ($M_\mathrm{thres}$) and with the approximate thermal treatment ($M_{\mathrm{\mathrm{thres}}}^{1.75}$) as a function of $M_\mathrm{thres}$. Hyperons can reduce $M_\mathrm{thres}$ by about 0.05~$M_\odot$ in comparison to the threshold mass determined from purely nucleonic matter with the same stellar parameters of cold NSs. As discussed in Ref.~\cite{Kochankovski:2025lqc}, we note that inferring the occurrence of hyperons would require a well measured $M_\mathrm{thres}$ together with the precise knowledge of the cold NS parameters to compare the measured threshold mass with the one expected for nucleonic EoSs.

\section{Conclusions}
We have reviewed the effects of hyperons in BNS mergers. Given the distinctive thermal behaviour of hyperons, we find that for hyperonic models the GW postmerger frequency is characteristically increased by a few per cent because of the reduced thermal pressure. This can be in principle used to discern hyperonic EoSs from purely nucleonic models. We also conclude that simulations with hyperonic EoSs feature systematically lower averaged temperatures in the remnant in line with the larger heat capacity of hyperonic matter. 
Moreover, mergers with hyperonic EoSs yield tentatively more ejecta than nucleonic models with a similar stellar radius. As for the threshold binary mass for prompt black-hole formation, we see that a hyperonic EoS can lead to a reduction of the threshold mass by about 0.05~$M_\odot$ in comparison to the purely nucleonic model with the same stellar parameters of cold NSs.

\section*{Acknowledgments}
This research has been supported from the projects CEX2020-001058-M, CEX2024-001451-M (Unidades de Excelencia ``Mar\'{\i}a de Maeztu"), PID2022-139427NB-I00 and PID2023-147112NB-C21 financed by MCIN/AEI/10.13039/501100011033/FEDER, UE. H.K. acknowledges support from the PRE2020-093558 Doctoral Grant of the spanish MCIN/ AEI/10.13039/501100011033/.  L.T. acknowledges support from the Generalitat Valenciana under contract CIPROM/2023/59 and from the CRC-TR 211 'Strong-interaction matter under extreme conditions'- project Nr. 315477589 - TRR 211. G.L acknowledges support by the Deutsche Forschungsgemeinschaft (DFG, German Research Foundation) - MA 4248/3-1 and support by the Klaus Tschira Foundation. S.B. and A.B. acknowledge support by Deutsche Forschungsgemeinschaft (DFG, German Research Foundation) through Project-ID 279384907 -- SFB 1245 (subproject B07). A.B. acknowledges support by the European Research Council (ERC) under the European Union’s Horizon 2020 research and innovation program under grant agreement No. 759253 and ERC Grant HEAVYMETAL No. 101071865 and support by the State of Hesse within the Cluster Project ELEMENTS.

%\begin{thebibliography}{99}
%\bibitem{...}
%\end{thebibliography}

\bibliographystyle{JHEP}
\bibliography{ref.bib}

\end{document}